\documentclass[namedate,webpdf,modern,medium]{oup-authoring-template}

\usepackage{graphicx}
\usepackage{booktabs}
\usepackage{amsmath}
\usepackage{amssymb}
\usepackage{hyperref}
\usepackage{xcolor}

\graphicspath{{Fig/}}
\renewcommand{\figurename}{Fig.}

\begin{document}

\journaltitle{Digital Scholarship in the Humanities}
\DOI{DOI to be assigned}
\copyrightyear{2026}
\pubyear{2026}
\vol{XX}
\issue{x}
\access{Submitted: 21 March 2026}
\appnotes{Original Article}

\title[AETAS for Legal History]{AETAS: Analysis of Evolving Temporal Affect and Semantics for Legal History}

\author[1,$\ast$]{Qizhi Wang}

\address[1]{\orgdiv{Data and AI-Innovation Lab}, \orgname{PingCAP}, \orgaddress{\state{CA}, \country{USA}}}

\corresp[$\ast$]{Corresponding author. \href{mailto:qizhi.wang@pingcap.com}{qizhi.wang@pingcap.com}}

\abstract{Digital-humanities work on semantic shift often alternates between handcrafted close readings and opaque embedding machinery. We present a reproducible expert-system style pipeline that quantifies lexical drift and its instability in the Old Bailey Corpus (1674--1913), coupling interpretable trajectories with legally meaningful axes. We bin proceedings by decade with dynamic merging for low-resource slices, train skip-gram embeddings, align spaces through orthogonal Procrustes to a 1900s anchor, and measure both geometric displacement and neighborhood turnover. We add split-half baselines and seed-sensitivity checks to separate within-bin instability from temporal change. Three visual analytics outputs (drift magnitudes, semantic trajectories, and movement along a mercy-versus-retribution axis) expose how \emph{justice}, \emph{crime}, \emph{poverty}, and \emph{insanity} evolve with penal reforms, transportation debates, and Victorian moral politics. The pipeline is implemented as auditable scripts so results can be reproduced in other historical corpora.}


\maketitle

\section{Introduction}
Semantic drift studies in digital humanities now leverage distributional semantics to explain cultural change \citep{hamilton2016,kutuzov2018}. Yet domain experts often face two gaps: (1) storylines that connect numeric drift to socio-legal events, and (2) tooling that behaves like an expert system with auditable steps and repeatable outputs. We address these gaps through \textbf{AETAS} (\textit{Analysis of Evolving Temporal Affect and Semantics}; Latin \textit{aetas}, meaning era or lifetime), a focused case study on the Old Bailey Corpus (OBC), a canonical source for London courtroom proceedings (1674--1913). The goal is to surface how legal and social concepts moved from early modern courts to late-Victorian professional justice.

A persistent obstacle is instability: embedding drift can reflect sampling variance or training noise as much as true semantic change. We therefore pair each drift claim with stability diagnostics (split-half baselines, seed sensitivity, and frequency controls), and report net drift/effect sizes that subtract within-bin instability before interpretation.

We contribute: (i) a reproducible pipeline that treats diachronic embeddings as chained inference steps, (ii) a stability-aware analysis that reports split-half baselines, seed sensitivity, anchor sensitivity, and net drift effect sizes, and (iii) a narrative with visual artefacts tuned for interpretability rather than black-box accuracy. Relative to prior SGNS+Procrustes pipelines, our emphasis is on stability diagnostics (aligned split-halves, permutation baselines), interpretability scaffolding (value axes with sensitivity and holdout checks), and explicit DH-ready exemplars. Compared to our earlier pipeline report, we add aligned split-half baselines with net-drift/effect sizes, anchor and seed sensitivity with chained alignment, hubness-aware overlap (CSLS) plus Global Anchors drift, permutation/downsampling controls, and axis sensitivity bands with a holdout sanity check.

\section{Related work}
Early semantic-change models used frequency- and context-based measures \citep{hamilton2016}. Subsequent surveys emphasise alignment stability, evaluation difficulties, and downstream interpretability \citep{kutuzov2018,tahmasebi2021}. Contextualised models and probing have been explored for finer-grained senses \citep{giulianelli2020}. Digital-humanities applications couple embeddings with historical argumentation, e.g., criminality, gender, and class \citep{hill2019,underwood2019}. Our work extends this line with an expert-system stance: each step (binning, alignment, drift scoring) is explicit, audited, and paired with narrative-ready visualisations.
Dynamic and joint-training approaches reduce alignment variance by learning shared temporal spaces, including dynamic embeddings \citep{bamler2017,rudolph2018}. We retain independently trained SGNS models for auditability and interpretability, but compare against chain alignment and alignment-free baselines to contextualise these alternatives.

\section{Data and methods}
\subsection{Corpus and temporal binning}
We use OBC 2.0 TEI-XML, converting proceedings to a parquet corpus covering 1674--1913. OBC 2.0 is human-transcribed rather than OCRed, but titles, numerals, and proper-noun noise persist; we quantify their impact via NER/POS and proxy filters. Text is lowercased, punctuation stripped except hyphens, and grouped into 10-year bins. Bins with fewer than five million tokens merge forward (resulting in an initial 1674s bin through 1749), preserving chronological order.

\subsection{Embedding training}
Skip-gram with negative sampling (vector size 300, window 5, min count 10, negative samples 10, epochs 5, seed 42) is trained per bin using \texttt{gensim} \citep{mikolov2013} with default subsampling ($t{=}10^{-3}$). Tokens per bin form sentence streams, avoiding shuffling across periods. We keep hyperparameters fixed across bins for comparability with standard diachronic SGNS settings; on 1{,}000-document samples, perturbations (window 2/8, dim 100, negative 5) preserve drift rankings (Spearman $\rho \in [0.52,0.60]$), so we avoid per-bin tuning in sparse slices. A no-subsampling check on 2{,}000-document samples preserves drift rankings (Spearman $\rho \in [0.94,1.00]$ across spans).

\subsection{Space alignment}
To compare meanings, we align each target space to a 1900s base using orthogonal Procrustes on the shared vocabulary \citep{hamilton2016}. Alignment uses the full shared vocabulary above the min-count threshold (10) in both bins; we do not cap the anchor set unless stated (e.g., Global Anchors). As a stable-anchor check, we select the 1{,}000 lowest-drift words among the top-20k shared types and re-align; drift rankings remain highly similar (Spearman $\rho{=}0.83$, $n{=}19$). Because these anchors are identified after an initial alignment, we treat the stable-anchor variant strictly as a robustness probe rather than a primary result. Let $X$ and $Y$ be base and target matrices; the rotation $R$ is obtained from $\text{SVD}(Y^\top X)=U\Sigma V^\top$ with $R=UV^\top$, ensuring distances are preserved.
Mean-centering the matrices before Procrustes yields near-identical drift rankings (Spearman $\rho$ in $[0.96,1.00]$ across spans, $n{=}6$--7), so we report uncentered results for consistency with prior work.

\subsection{Drift scoring and neighbors}
Semantic drift between a word vector $v_t$ at start and $v_{t'}$ at end is $d=1-\cos(v_t, v_{t'})$. Neighbors are computed in the aligned spaces using the full vocabulary for each bin (no re-normalization is needed after orthogonal rotation). We track neighborhood turnover via Jaccard overlap between top-$k$ neighbors at the two times, $J=\frac{|N_t \cap N_{t'}|}{|N_t \cup N_{t'}|}$.

\subsection{Stability baselines and net drift}
To quantify within-bin instability, we perform split-half training: for each bin, we randomly split texts into halves, train SGNS on each half (fixed seed to isolate sampling variance), align the two half spaces with orthogonal Procrustes, and measure within-bin drift for focal words. We repeat this 20 times and report mean and standard deviation; seed variance is checked separately via multi-seed incremental SGNS. For a cross-bin span $(t,t')$, we define net drift as
\[
d_{\text{net}} = d_{t,t'} - \frac{s_t + s_{t'}}{2},
\]
where $s_t$ and $s_{t'}$ are split-half means for the start and end bins. We also report a standardized effect size
\[
z = \frac{d_{\text{net}}}{(\sigma_t + \sigma_{t'})/2},
\]
using the split-half standard deviations $\sigma_t, \sigma_{t'}$ where available. Net drift is interpreted as change that exceeds within-bin instability. We treat $z$ as a diagnostic rather than a significance test and additionally report seed-bootstrap uncertainty for drift (5 seeds on 2{,}000-document samples; median 95\% CI width 0.054 across 19 word-span pairs).
Table~\ref{tab:netdrift} lists net drift and $z$ values for all focal terms across the main spans.

\subsection{Normative axis}
To ground interpretability, we build a mercy-retribution axis using expanded seed sets of \{\textit{mercy, pity, charity, kindness, lenity, forgiveness, clemency}\} minus explicitly punitive terms \{\textit{punishment, penalty, execution, imprisonment, sentence, hanging, severity}\}, normalized to unit length. The axis is computed per time slice from that slice's vectors. Projections of target terms on this axis (cosine to the axis) capture value-laden movement beyond raw drift. A holdout sanity check using non-seed mercy lexemes (e.g., \textit{pardon, grace, charitable, merciful}) versus punitive lexemes (e.g., \textit{gaol, prison, conviction, discipline}) is reported in the robustness section.

\subsection{Reproducibility}
All stages (corpus parsing, binning, SGNS training, alignment, drift scoring, figure generation) are scripted for repeatable execution on other corpora; the repository includes \texttt{run-all.sh} and a frozen environment spec (\texttt{uv.lock}, Python 3.12, gensim 4.3, transformers 4.57) with fixed random seeds. The repository URL and commit hash will be released at camera-ready; artifacts are available upon request. Because OBC is distributed under its own license, we do not redistribute raw TEI text; we will release derived drift tables, figures, and scripts to reconstruct results for licensed users. A minimal synthetic MRE (script + toy corpus) that runs end-to-end is included in the repository and can be shared during review upon request.

\subsection{Corpus diagnostics}
Table~\ref{tab:corpus} reports decadal statistics. Early sparse decades (1674--1749) merge to satisfy the 5M-token rule; peak coverage is 1840s (13.2M tokens, 85K types). Headline drift spans begin at 1750s, with the early merged bin used only for chain-alignment diagnostics. A naive lemmatisation heuristic reduces vocabulary size by 11--15\% across decades; a conservative orthography normalization on sampled bins reduces vocabulary by 6--7\%, indicating moderate fragmentation without dominating the counts.

\begin{table}[htbp]
  \centering
  \small
  \caption{Corpus size by decade after adaptive merging. Tokens are raw token counts; vocab is type count after lowercasing and punctuation stripping.}
  \label{tab:corpus}
  \begin{tabular}{lrrr}
    \toprule
    Bin & Years & Tokens (M) & Vocab (K) \\
    \midrule
    1674s & 1674--1749 & 5.72 & 56.6 \\
    1750s & 1750--1769 & 5.16 & 40.6 \\
    1770s & 1770--1789 & 9.51 & 51.1 \\
    1790s & 1790--1799 & 6.09 & 42.3 \\
    1800s & 1800--1809 & 5.67 & 40.5 \\
    1810s & 1810--1819 & 5.60 & 46.6 \\
    1820s & 1820--1829 & 8.40 & 54.4 \\
    1830s & 1830--1839 & 12.38 & 76.9 \\
    1840s & 1840--1849 & 13.18 & 85.1 \\
    1850s & 1850--1859 & 9.99 & 59.0 \\
    1860s & 1860--1869 & 8.79 & 59.5 \\
    1870s & 1870--1879 & 7.98 & 58.5 \\
    1880s & 1880--1889 & 9.57 & 60.4 \\
    1890s & 1890--1899 & 7.46 & 55.1 \\
    1900s & 1900--1909 & 8.41 & 57.6 \\
    1910s & 1910--1913 & 2.59 & 32.5 \\
    \bottomrule
  \end{tabular}
\end{table}

\section{Results}
\subsection{Drift magnitudes}
\begin{figure*}[t]
  \centering
  \includegraphics[width=0.98\textwidth]{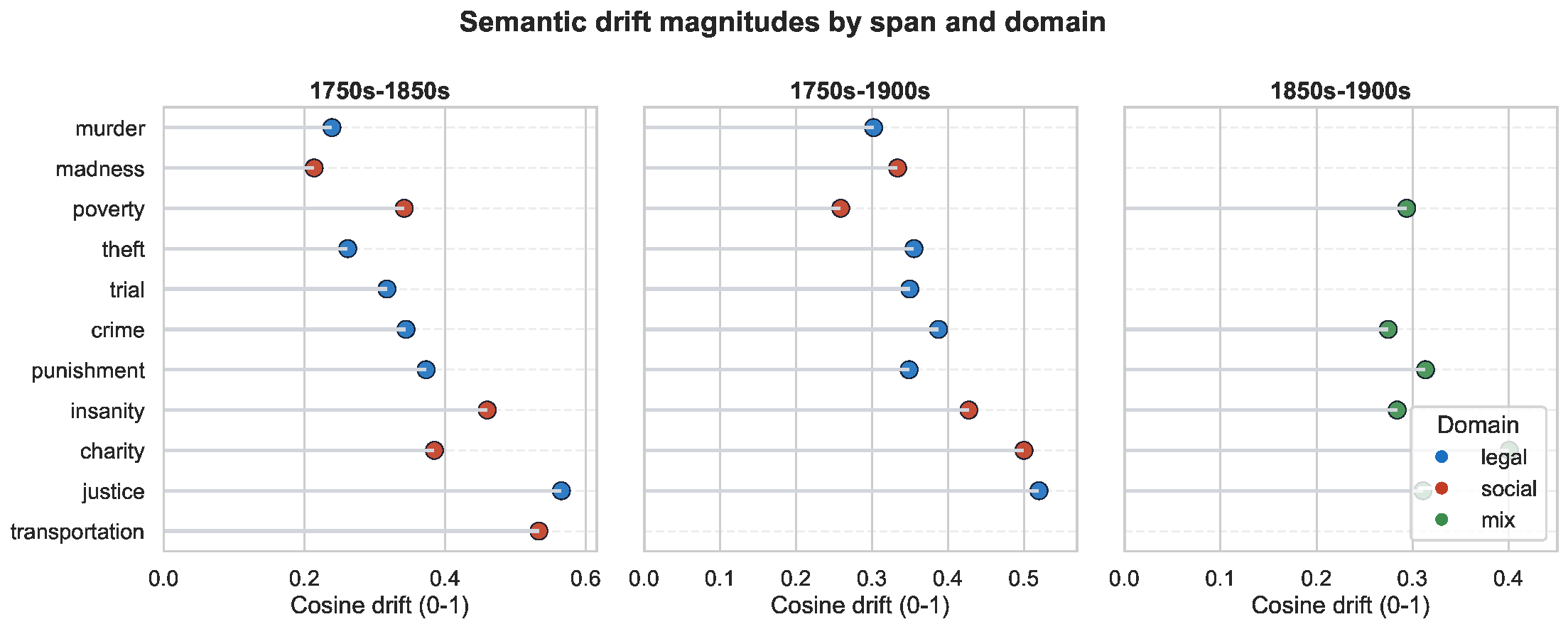}
  \caption{Semantic drift magnitudes by span and domain. Points show cosine drift; stems indicate displacement relative to each time window.
\noindent\textit{Alt text: Lollipop chart of cosine drift values for focal terms across spans; larger dots indicate greater drift.}}
  \label{fig:lollipop}
\end{figure*}

Fig.~\ref{fig:lollipop} shows cosine drift per target word across spans. On average, drift is $0.36$, with the largest shifts in \textit{justice} (0.57 in 1750s--1850s) and \textit{transportation} (0.53), reflecting the decline of deportation and the rise of professional judges. Social terms such as \textit{charity} exhibit long-horizon drift (0.50) as philanthropy professionalises.

Table~\ref{tab:topdrift} summarises the largest displacements and sparse neighbor overlap, highlighting that shifts are driven by contextual replacement rather than stable colexification.

To broaden the target set, we computed drift for the top 500 shared high-frequency words per span (frequency-controlled by minimum count across bins). The focal terms occupy high percentiles in this larger distribution: \textit{justice} and \textit{transportation} fall in the 99--100th percentile for 1750s--1850s, \textit{charity} and \textit{justice} are above the 98th percentile for 1750s--1900s, and most focal words sit above the 70th percentile for 1850s--1900s. This suggests the highlighted shifts are not artifacts of a tiny curated list, even though some targets (e.g., \textit{poverty} in 1750s--1900s) are closer to the median. The full shared vocabularies contain 7{,}236 (1750s--1850s), 6{,}628 (1750s--1900s), and 10{,}446 (1850s--1900s) types above the min-count threshold; the repeated $n{=}19$ correlations correspond to seven focal words across three spans, with \textit{transportation} falling below min-count in the 1900s (21 possible pairs $\rightarrow$ 19 observed). Thus \textit{transportation} is only reported for 1750s--1850s, and is excluded from 1750s--1900s/1850s--1900s tables and figures. To support independent scrutiny, we also export full ranked lists for the top-1{,}000 shared words per span, with frequency and sampled POS estimates. Windowed cue analysis around \textit{transportation} indicates penal cues dominate while logistical cues are rare (penal-cue shares 0.07 in 1750s and 0.73 in 1850s; logistical cues $<0.004$), and the term’s frequency collapses from 2{,}283 occurrences in 1750s to 62 in 1850s and 3 in 1900s, suggesting the drift reflects the disappearance of the penal sense rather than a shift to logistics.

\begin{table}[t]
  \centering
  \small
  \setlength{\tabcolsep}{3pt}
  \caption{Top drift cases across spans. Overlap is the Jaccard of top-5 neighbors at the start and end periods.}
  \label{tab:topdrift}
  \begin{tabular}{llllr}
    \toprule
    Word & Span & Domain & Drift & Overlap \\
    \midrule
    justice & 1750s--1850s & legal & 0.565 & 0.000 \\
    transportation & 1750s--1850s & social & 0.533 & 0.000 \\
    justice & 1750s--1900s & legal & 0.520 & 0.000 \\
    charity & 1750s--1900s & social & 0.500 & 0.000 \\
    insanity & 1750s--1850s & social & 0.460 & 0.000 \\
    insanity & 1750s--1900s & social & 0.427 & 0.000 \\
    charity & 1850s--1900s & mix & 0.401 & 0.000 \\
    crime & 1750s--1900s & legal & 0.388 & 0.111 \\
    \bottomrule
  \end{tabular}
\end{table}

\subsection{Semantic trajectories}
\begin{figure}[htbp]
  \centering
  \includegraphics[width=0.95\columnwidth]{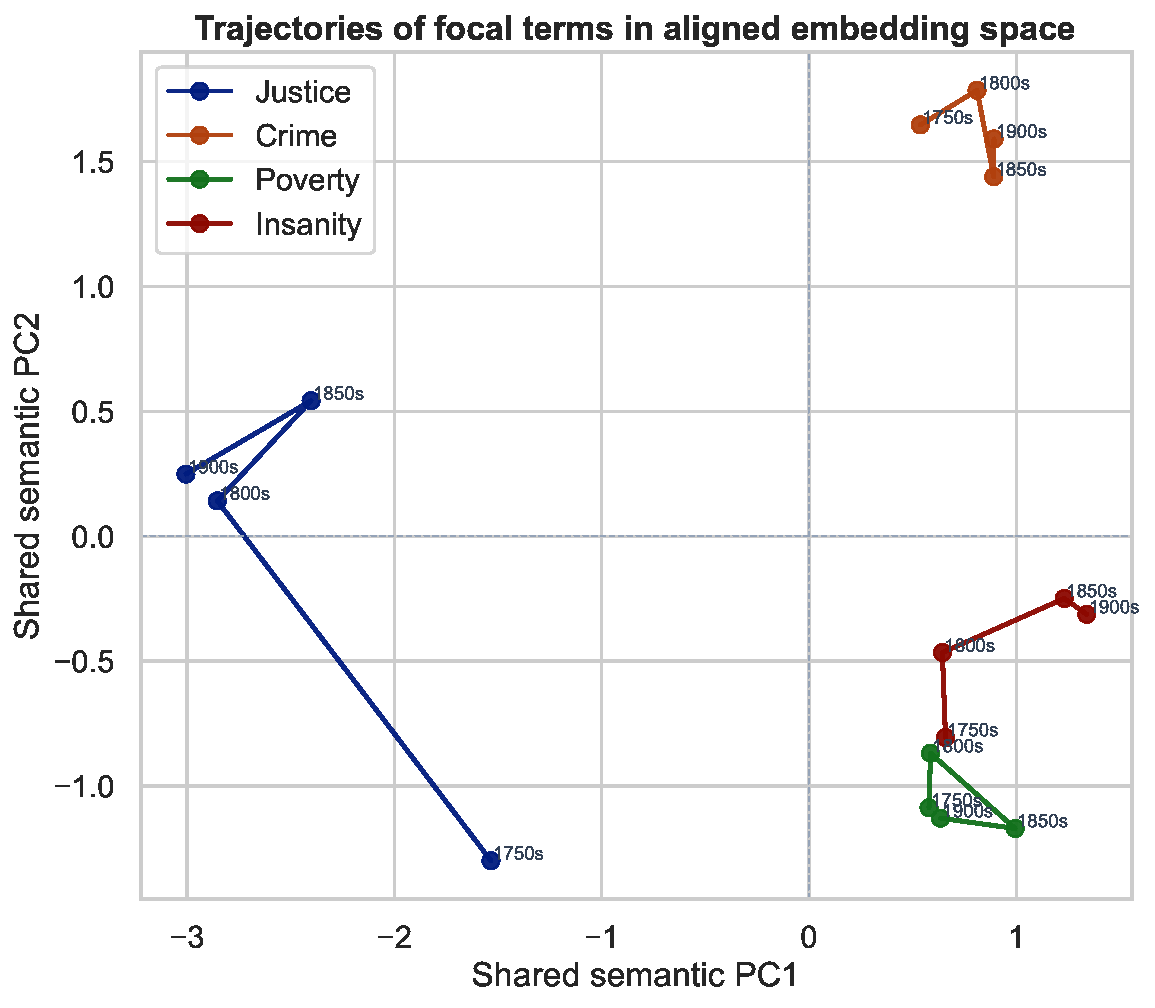}
  \caption{Aligned semantic trajectories for focal terms. Arrows indicate chronological order toward 1900s.
\noindent\textit{Alt text: PCA trajectory lines for focal terms across decades with arrows pointing toward the 1900s positions.}}
  \label{fig:trajectory}
\end{figure}

Fig.~\ref{fig:trajectory} plots PCA projections of aligned vectors. \textit{Justice} moves from a dispersed early-modern sense (neighbors include \textit{baron}, \textit{recorder}) toward a tight legal-professional cluster (\textit{bigham}, \textit{phillimore}). \textit{Poverty} shifts away from moral failing toward structural conditions (neighbors like \textit{ill-health}, \textit{schemes} in 1900s).

\subsection{Close-reading exemplars}
To ground the quantitative narrative, we sampled short excerpts from the corpus. In the 1750s bin, \textit{justice} appears as an abstract goal: ``\ldots brought to justice, that I may be righted.'' In the 1900s bin, it appears as a judicial title: ``(BEFORE MR. JUSTICE BRAY.)'' For \textit{poverty} in the 1850s, mitigation language is explicit: ``Recommended to mercy, in consequence of her poverty.'' For \textit{insanity} in the 1900s, the term is framed medically: ``\ldots in a state of distraction bordering on insanity, as the only way of ventilating his wrongs.'' These snippets align with the measured shift toward professionalised court roles and medicalised explanations.

\subsection{Value-laden movement}
\begin{figure}[htbp]
  \centering
  \includegraphics[width=0.95\columnwidth]{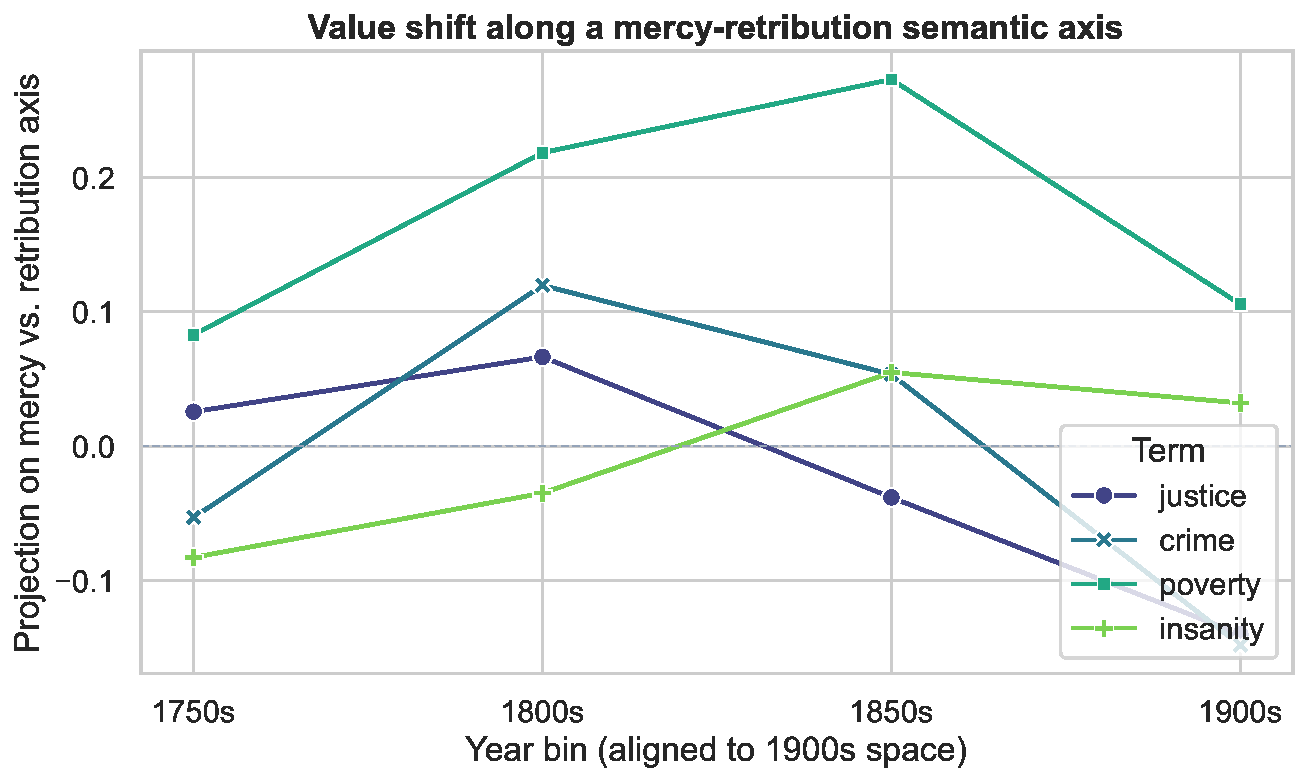}
  \caption{Shifts along a mercy-retribution axis. Positive scores indicate proximity to mercy-related terms; negatives lean toward punitive semantics.
\noindent\textit{Alt text: Line plot of mercy–retribution axis projections for each focal word over time; justice is negative by 1900s.}}
  \label{fig:normative}
\end{figure}

Fig.~\ref{fig:normative} shows projections on the mercy-retribution axis. \textit{Justice} drifts slightly toward a more punitive neighbourhood by 1900s (score $-0.14$), consistent with late-Victorian emphasis on deterrence despite procedural reforms. \textit{Insanity} and \textit{poverty} trend toward the merciful side by 1850s before softening in 1900s, echoing medicalisation and social-reform rhetoric. Fig.~\ref{fig:sensitivity} adds seed-variation bands; the sign of \textit{justice} is robust to leave-one-out seeds, while \textit{poverty} shows wider sensitivity because it co-occurs with both moralising and structural contexts. Using a fixed axis defined in the 1900s and projecting earlier bins via alignment yields similar scores and a strong correlation with the per-slice axis (Spearman $\rho{=}0.77$, $n{=}20$), suggesting the axis is not purely driven by seed drift. An antonym-pair axis (mercy--punishment, clemency--execution, lenity--severity, etc.) yields the same negative sign for \textit{justice} in the 1900s ($-0.24$). A medicalisation--moralisation axis (medical vs. moral seeds) shifts \textit{insanity} from negative in the 1750s ($-0.09$) to positive by the 1850s/1900s ($\approx$0.26), consistent with the narrative.

\subsection{Robustness checks}
\begin{figure}[htbp]
  \centering
  \includegraphics[width=0.95\columnwidth]{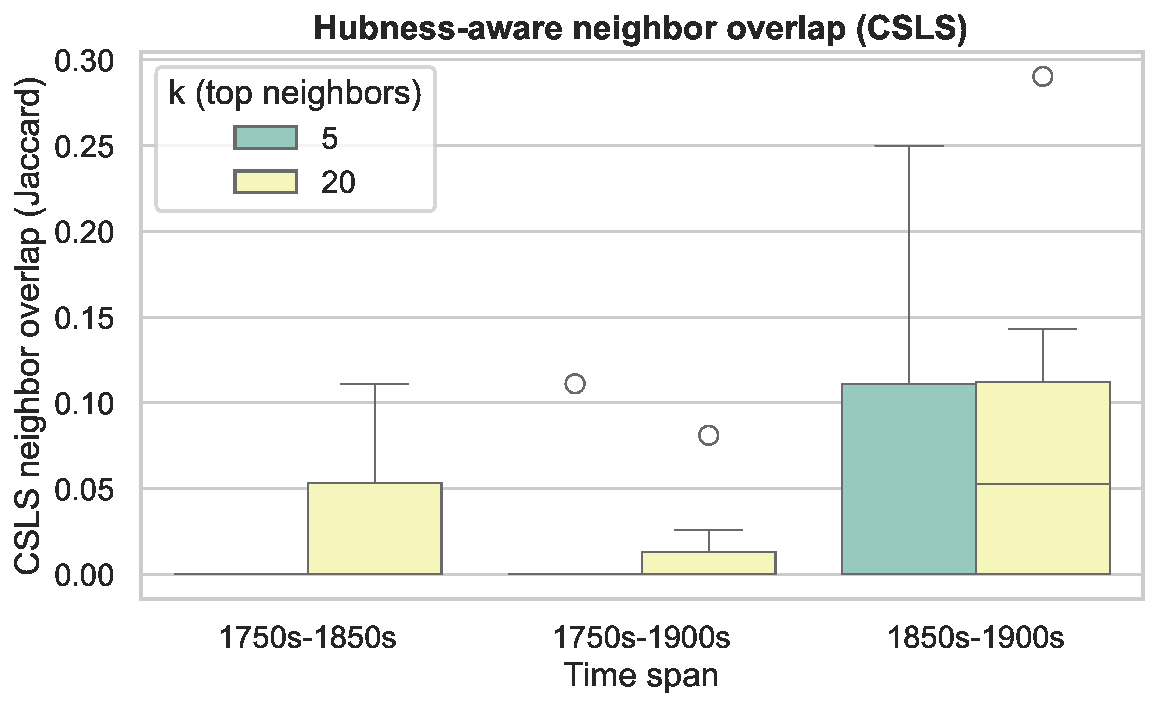}
  \caption{Hubness-aware neighbor overlap using CSLS on a capped 10k vocabulary.
\noindent\textit{Alt text: Bar chart of CSLS-based neighbor overlap for focal terms on a 10k vocabulary.}}
  \label{fig:csls}
\end{figure}

To mitigate brittleness, we enlarged neighborhoods to $k{=}20$ (Fig.~\ref{fig:overlap}). Jaccard overlap rises modestly (mean 0.044 vs. 0.032 for $k{=}5$), indicating that low overlap is not solely a small-$k$ artifact. Using hubness-aware CSLS neighbors on a capped 10k vocabulary yields similarly low but slightly higher overlap (mean 0.028 for $k{=}5$, 0.042 for $k{=}20$; Fig.~\ref{fig:csls}). Rank-biased overlap (RBO, $p{=}0.9$) remains low (mean 0.047), and POS-restricted neighbors (NOUN/VERB/ADJ) increase slightly (mean 0.054), suggesting low overlap reflects genuine reconfiguration rather than only hubness or named entities. A qualitative audit of top neighbors clarifies the zero-Jaccard cases: in the 1750s, \textit{justice} neighbors include \textit{magistrate}, \textit{roundhouse}, and \textit{deposition}, whereas in the 1900s they are dominated by judge surnames such as \textit{bigham}, \textit{grantham}, and \textit{coleridge}, indicating replacement rather than hubness artifacts. Removing judge surnames from neighbor lists leaves \textit{justice} with zero Jaccard overlap across all spans, and other focal terms remain low ($\leq 0.18$), so near-zero overlap is not driven solely by named-entity carryover. Anchor sensitivity tests show high agreement between alternative anchors and the 1900s anchor (Spearman $\rho{=}0.97$ for 1850s, $\rho{=}0.95$ for 1880s; 19 word-span pairs). A sequential chain alignment (1674s$\rightarrow$1910s) produces drift rankings consistent with the 1900s anchor (Spearman $\rho{=}0.88$ over 19 word-span pairs); starting the chain at 1750s (excluding the early mega-bin) yields essentially the same agreement (Spearman $\rho{=}0.88$, $n{=}19$). Stable-anchor alignment (1{,}000 lowest-drift anchors among the top-20k shared types) yields similar rankings (Spearman $\rho{=}0.83$, $n{=}19$). Downsampling 1850s/1900s bins to the 1750s token count yields moderate rank agreement with full-data drifts (Spearman $\rho{=}0.57$ for 1750s--1850s, $\rho{=}0.80$ for 1750s--1900s), suggesting that bin-size imbalance affects magnitude but not the highest-level ordering. Raising the merge threshold to 7M shifts early bins (start $\rightarrow$ 1760s), but the 1850s--1900s drift ranking in sampled models matches the 5M baseline for overlapping terms (Spearman $\rho{=}1.0$, $n{=}5$). On 1{,}000-document per-bin samples, hyperparameter perturbations (window 2/8, dim 100, negative 5) preserve rankings with Spearman $\rho$ in $[0.52, 0.60]$ ($n{=}9$). A coarse change-point scan over 1750s$\rightarrow$1810s$\rightarrow$1850s$\rightarrow$1900s shows the largest drift for \textit{justice}, \textit{transportation}, and \textit{charity} in 1750s--1810s, while \textit{poverty} and \textit{insanity} peak in 1810s--1850s, aligning with early nineteenth-century reforms. A case-aware proper-noun proxy (dropping capitalized non-initial tokens, numerals, and honorific patterns) run on the main bins yields drift rankings highly correlated with the baseline (Spearman $\rho{=}0.91$, $n{=}15$) and keeps \textit{justice}/\textit{transportation} among the highest drifts; mean overlap becomes 0.037 for $k{=}5$ and 0.040 for $k{=}20$, so low overlap is not solely a named-entity artifact. A corpus-wide title filter shows \textit{justice} tokens preceded by titles rise from 0.7\% in the 1750s to 77--80\% in the 1850s/1900s; training on the filtered corpus keeps drift large (0.51 for 1750s--1850s; 0.57 for 1750s--1900s) and retains a negative 1900s axis projection ($-0.16$), indicating the shift is not solely an honorific artifact. Using TEI judiciary-name lists to drop \textit{justice} tokens followed by judge surnames removes 0.5\%, 7.2\%, and 46.7\% of \textit{justice} tokens in 1750s/1850s/1900s; drift remains high (0.61 for 1750s--1850s; 0.53 for 1750s--1900s), and the 1900s axis projection stays negative ($-0.19$). Seed-bootstrap CIs (5 seeds on 2{,}000-document samples) yield a median 95\% CI width of 0.054 across 19 word-span pairs, indicating moderate uncertainty without overturning the main ordering. Time-shuffle permutation tests over all targets/spans yield $p_{\ge}$ values in $[0.18, 1.00]$, with \textit{justice} above the null mean (1.32--1.53 SDs for 1750s--1850s/1900s) and \textit{transportation} modestly above null (0.55 SDs in 1750s--1850s); we treat permutation results as a conservative sanity check rather than a significance claim. A frequency-matched random-seed null for the 1900s axis shows most target projections are not extreme (e.g., \textit{justice} $p_{\ge}{=}0.13$), while \textit{charity} is unusually high ($p_{\ge}{=}0.005$). Target-word frequencies per million tokens stay within the same order across bins (e.g., \textit{justice} 682$\rightarrow$90), suggesting drift is not driven by single bursts. Conservative lemmatisation/orthography normalization reduces type counts by 6--15\%, implying that spelling variation inflates vocabulary moderately but does not dominate. A holdout axis sanity check finds non-seed mercy lexemes scoring higher than punitive lexemes in the 1900s bin (mean 0.07 vs. $-0.13$, $n{=}7$ vs. 12).

\subsection{Contextual, CCA, VecMap, pivot, and Global Anchors baselines}
We ran a contextual BERT-base baseline (up to 120 contexts/bin; early bins are sparse) with APD over contextual embeddings and substitute-based scaled JSD. Transportation shows the largest APD and JSD in 1750s--1850s, mirroring the SGNS signal; \textit{justice} APD is moderate. A CCA alignment baseline (50 components) yields drift for \textit{justice} and \textit{transportation} that aligns with SGNS trends, with mean canonical correlations 0.63--0.68 over 6k--10k shared types. VecMap unsupervised alignment (capped vocabulary) produces consistent rankings (e.g., \textit{justice} high drift in 1750s--1850s). To avoid explicit alignment, we compute a pivot baseline: rank similarity to 500 high-frequency pivots per bin and compare top-50 lists; pivot Jaccard remains low for \textit{justice} but higher for \textit{crime}/\textit{poverty}, with mid-level Spearman stability. A temporal-reference norm baseline (anchor 1900s) also shows large norms for \textit{justice} and \textit{transportation}, consistent with long-range drift. As an alignment-free global measure, Global Anchors (1,000 shared anchors) produces drift rankings correlated with SGNS (Spearman $\rho{=}0.72$, $n{=}19$), again highlighting \textit{justice} and \textit{transportation} as high-shift terms. As an interpretable alternative, a dependency-slot JSD baseline on sampled contexts yields high divergence for \textit{justice} and \textit{transportation} in 1750s--1850s (JSD $\approx$0.65--0.69 where counts are sufficient), consistent with the embedding signal. Across spans, SGNS rank correlations with baselines are: Global Anchors $\rho$ in $[0.66,0.75]$ ($n{=}6$--7), CCA $\rho$ in $[0.14,0.94]$ ($n{=}6$--7), APD/JSD $\rho$ in $[0.50,1.00]$ where contextual counts permit ($n{=}3$--5), and VecMap $\rho{=}1.00$ on the small span where it converged ($n{=}3$). Pivot similarity is inversely related to drift (Spearman $\rho$ in $[-0.96,-0.83]$), as expected since lower Jaccard indicates greater change.

\subsection{Incremental SGNS and multi-seed variance}
We trained an incremental (time-forward) SGNS with three seeds over 1750s$\rightarrow$1850s$\rightarrow$1900s (vector size 200, 3 epochs per bin), saving snapshots per bin. Drift means (std across seeds) are: \textit{justice} 0.40 (0.03) for 1750s--1850s, 0.50 (0.02) for 1750s--1900s; \textit{transportation} 0.14 (0.01) for 1750s--1850s; \textit{crime} 0.30 (0.01) for 1750s--1900s. Variance is low (std $\leq 0.03$) indicating seed robustness; this incremental model provides a simple dynamic baseline and preserves the main drift ordering for \textit{justice} and \textit{transportation}.

\subsection{Split-half stability}
\begin{figure}[htbp]
  \centering
  \includegraphics[width=0.95\columnwidth]{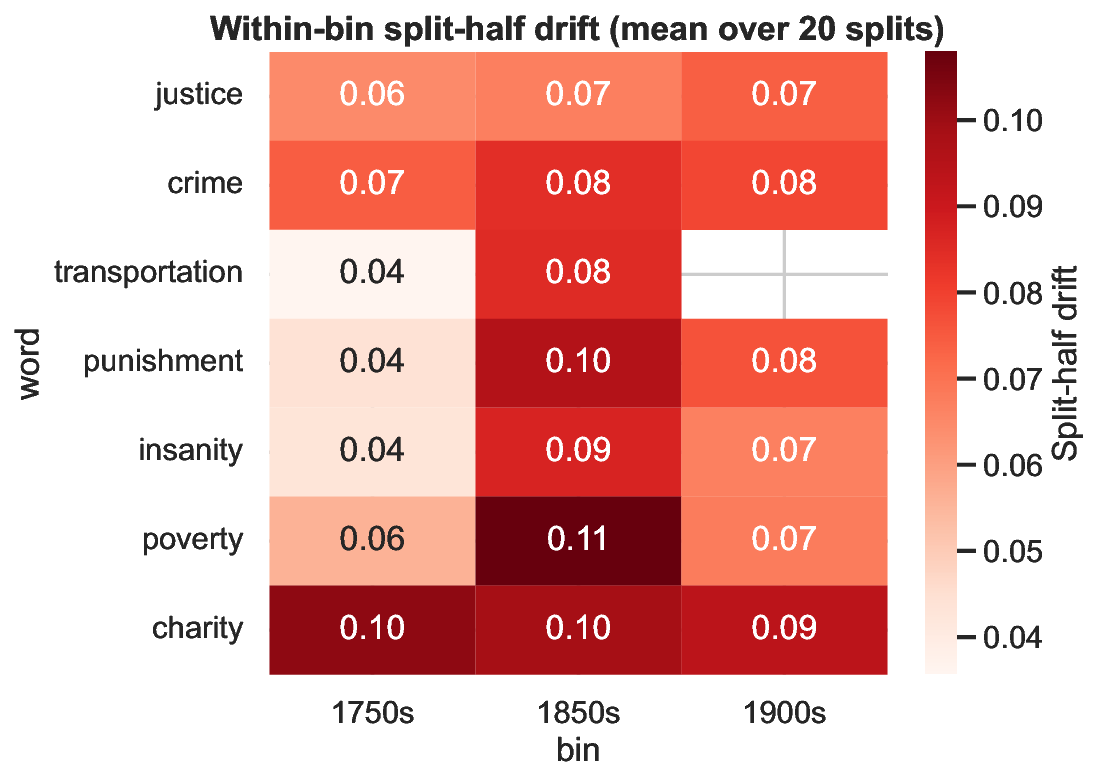}
  \caption{Split-half drift (mean over 20 splits).
\noindent\textit{Alt text: Heatmap of split-half within-bin drift values across words and bins.}}
  \label{fig:splithalf}
\end{figure}

Splitting each bin (1750s, 1850s, 1900s) into halves, training SGNS on each half (20 repeats), and aligning the half models before comparison yields modest within-bin drift (bin-level means $\approx$0.06, 0.09, 0.08 respectively; Fig.~\ref{fig:splithalf}). Net drift after subtracting these baselines remains positive across target terms, with the largest net shifts in \textit{justice} and \textit{transportation} (Fig.~\ref{fig:netdrift}). Standardized effect sizes are large because split-half variance is small; we therefore interpret $z$ as a stability diagnostic rather than a strict statistical test. Table~\ref{tab:netdrift} provides the full net-drift and $z$ values. Using full-corpus split-half variance, per-target 95\% intervals for net drift remain positive for all focal words (e.g., \textit{justice} 0.478--0.521 for 1750s--1850s), and are reported alongside released artifacts.

\subsection{Frequency controls and preprocessing}
\begin{figure}[htbp]
  \centering
  \includegraphics[width=0.95\columnwidth]{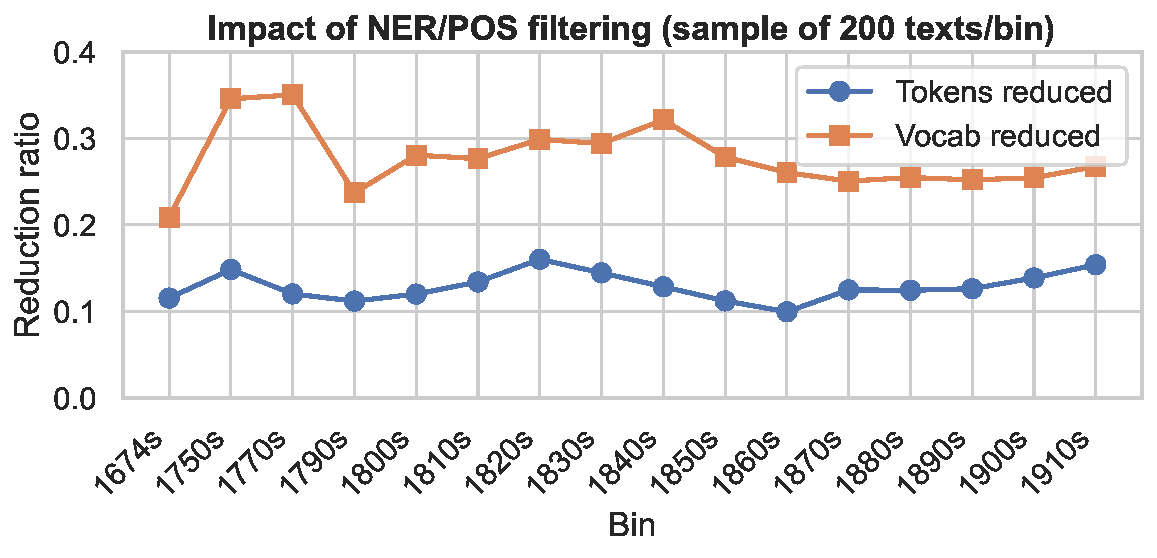}
  \caption{NER/POS filtering reductions.
\noindent\textit{Alt text: Bar chart showing percent token and type reduction after NER and POS filtering.}}
  \label{fig:nerfilter}
\end{figure}

Regressing SGNS drift on mean and delta frequency across spans yields no significant coefficients (p$>$0.09) on a small sample (27 word-span pairs; $R^2{=}0.54$), so we treat frequency controls as descriptive rather than confirmatory; contextual APD and incremental SGNS show similarly non-significant frequency effects (p$>$0.14, $R^2{<}0.44$). Partial correlations are weak: drift vs. log mean controlling log delta yields $r{=}-0.09$, while drift vs. log delta controlling log mean yields $r{=}0.34$ ($n{=}19$). A lightweight NER/POS filter on sampled texts removes 10--16\% of tokens and 21--35\% of types (Fig.~\ref{fig:nerfilter}), indicating named entities and numerals contribute noticeable noise. We additionally run a full case-aware proper-noun proxy (capitalized non-initial tokens, numerals, honorific patterns) on the main bins and report its drift/overlap stability in the robustness section. A conservative orthography normalization (plural-only lemmatisation plus British/American variant mapping) reduces vocabulary by 6--7\% on 2{,}000-document samples; drift rankings correlate strongly with the unnormalized sample (Spearman $\rho{=}0.92$, $n{=}13$), and \textit{justice}/\textit{transportation} remain the highest-drift terms among those retained. We therefore treat normalization as a qualitative check rather than a replacement for the main pipeline. Preprocessing for the baseline runs: lowercase, strip punctuation except hyphens, no stopword removal, no lemmatization.
\subsection{NER/POS filtering sanity check}
Applying a lightweight NER/POS filter (spaCy \texttt{en\_core\allowbreak\_web\allowbreak\_sm}) to 200-sample texts per bin removes 10--16\% of tokens and 21--35\% of types, indicating non-trivial proper-noun/number noise. To scale to the full corpus, we run a case-aware proxy filter (dropping capitalised non-initial tokens, numerals, and honorific patterns) on the main bins and rerun the pipeline. Drift rankings remain highly correlated with the baseline (Spearman $\rho{=}0.91$, $n{=}15$), with slight overlap gains (mean Jaccard 0.037 for $k{=}5$, 0.040 for $k{=}20$), suggesting that named-entity noise is not the primary driver. Sampled POS tagging of occurrences shows \textit{justice} shifting toward PROPN usage (PROPN share 0.56 in 1750s, 1.00 in 1850s, 0.91 in 1900s), consistent with the title-driven shift.

\begin{figure}[htbp]
  \centering
  \includegraphics[width=0.95\columnwidth]{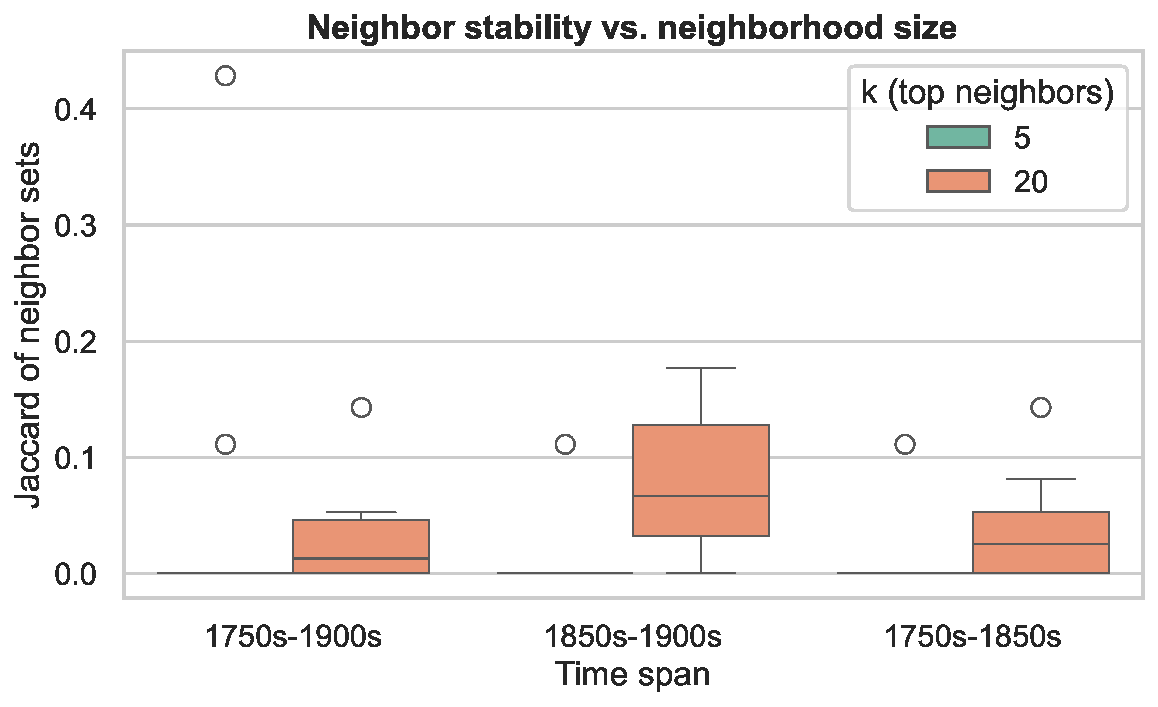}
  \caption{Neighbor stability when increasing $k$ from 5 to 20 (Jaccard of start/end neighborhoods). Larger $k$ raises overlap but many terms still shift into new contextual neighborhoods.
\noindent\textit{Alt text: Bar chart comparing Jaccard neighbor overlap for $k{=}5$ versus $k{=}20$ across focal terms.}}
  \label{fig:overlap}
\end{figure}

\begin{figure}[htbp]
  \centering
  \includegraphics[width=0.95\columnwidth]{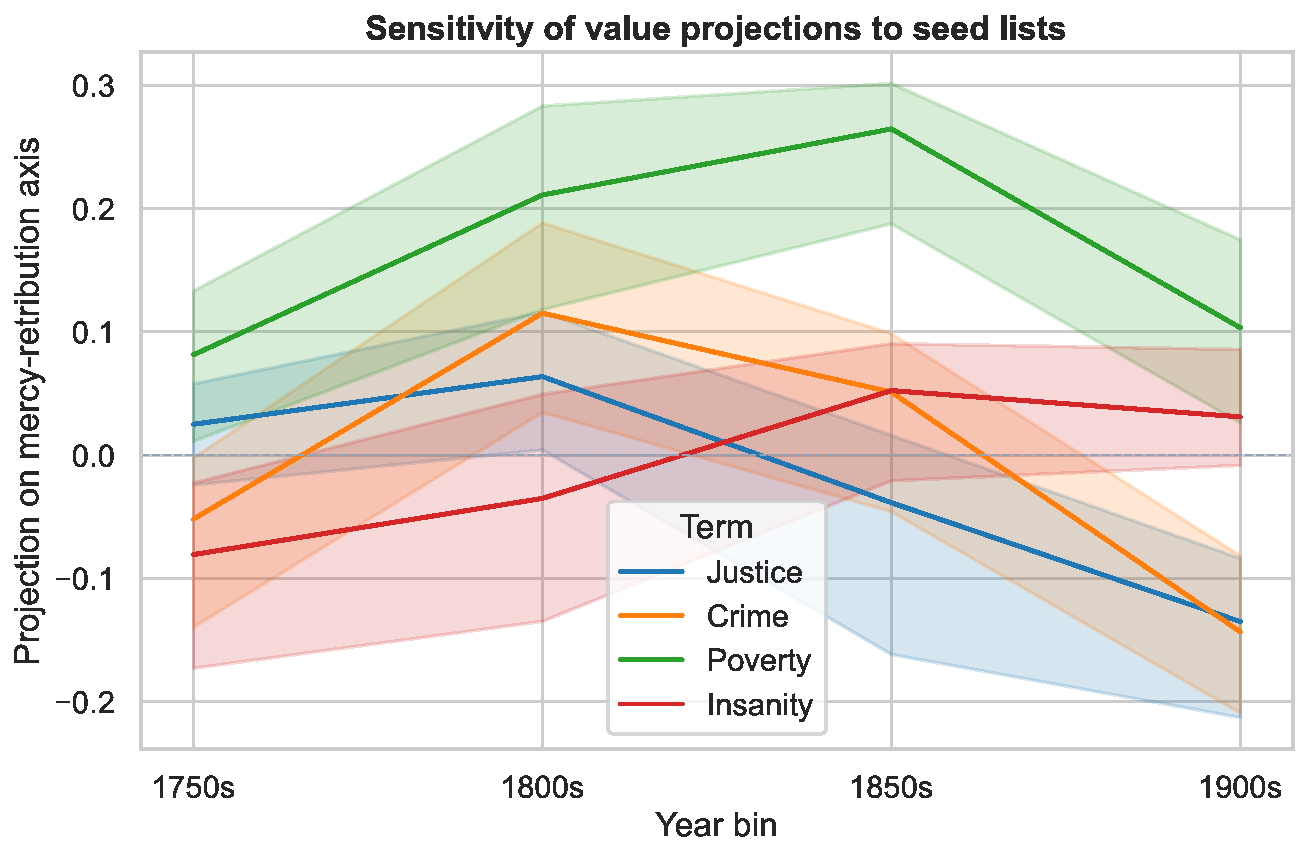}
  \caption{Mercy-retribution projections under leave-one-out seed variants. Bands span min/max per year; lines show mean. \textit{Justice} remains negative in 1900s across seed choices.
\noindent\textit{Alt text: Uncertainty bands showing min–max axis projections under leave-one-out seeds for each focal word.}}
  \label{fig:sensitivity}
\end{figure}

\begin{figure*}[t]
  \centering
  \includegraphics[width=\textwidth]{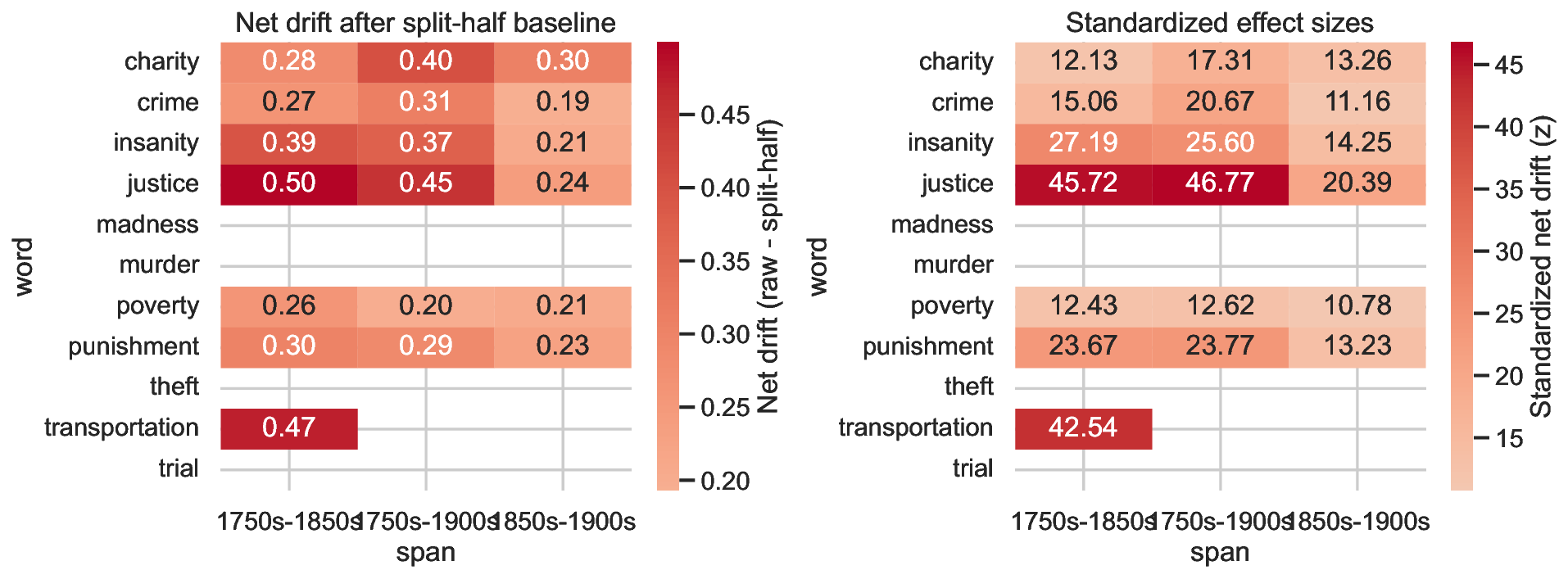}
  \caption{Net drift after subtracting split-half baselines (left) and standardized effect sizes (right); positive values mean drift exceeds within-bin instability.
\noindent\textit{Alt text: Side-by-side bars of net drift and standardized effect size per term and span.}}
  \label{fig:netdrift}
\end{figure*}

\begin{table}[t]
  \centering
  \small
  \setlength{\tabcolsep}{3pt}
  \caption{Net drift and effect sizes for focal terms across the main spans. ``Split'' is the mean split-half baseline; ``Net'' subtracts the start/end baselines. ``--'' indicates the term falls below the min-count threshold in the later bin (e.g., \textit{transportation} in the 1900s).}
  \label{tab:netdrift}
  \begin{tabular}{llrrrr}
    \toprule
    Word & Span & Drift & Split & Net & $z$ \\
    \midrule
    \textit{justice} & 1750s-1850s & 0.565 & 0.066 & 0.499 & 45.7 \\
    \textit{justice} & 1750s-1900s & 0.520 & 0.069 & 0.450 & 46.8 \\
    \textit{justice} & 1850s-1900s & 0.311 & 0.071 & 0.240 & 20.4 \\
    \textit{crime} & 1750s-1850s & 0.344 & 0.079 & 0.265 & 15.1 \\
    \textit{crime} & 1750s-1900s & 0.388 & 0.077 & 0.311 & 20.7 \\
    \textit{crime} & 1850s-1900s & 0.275 & 0.081 & 0.193 & 11.2 \\
    \textit{poverty} & 1750s-1850s & 0.342 & 0.082 & 0.260 & 12.4 \\
    \textit{poverty} & 1750s-1900s & 0.259 & 0.062 & 0.197 & 12.6 \\
    \textit{poverty} & 1850s-1900s & 0.294 & 0.088 & 0.206 & 10.8 \\
    \textit{insanity} & 1750s-1850s & 0.460 & 0.065 & 0.395 & 27.2 \\
    \textit{insanity} & 1750s-1900s & 0.427 & 0.055 & 0.373 & 25.6 \\
    \textit{insanity} & 1850s-1900s & 0.284 & 0.077 & 0.207 & 14.2 \\
    \textit{charity} & 1750s-1850s & 0.385 & 0.100 & 0.284 & 12.1 \\
    \textit{charity} & 1750s-1900s & 0.500 & 0.098 & 0.402 & 17.3 \\
    \textit{charity} & 1850s-1900s & 0.401 & 0.096 & 0.305 & 13.3 \\
    \textit{transportation} & 1750s-1850s & 0.533 & 0.060 & 0.473 & 42.5 \\
    \textit{transportation} & 1750s-1900s & -- & -- & -- & -- \\
    \textit{transportation} & 1850s-1900s & -- & -- & -- & -- \\
    \textit{punishment} & 1750s-1850s & 0.373 & 0.070 & 0.303 & 23.7 \\
    \textit{punishment} & 1750s-1900s & 0.349 & 0.060 & 0.289 & 23.8 \\
    \textit{punishment} & 1850s-1900s & 0.313 & 0.086 & 0.228 & 13.2 \\
    \bottomrule
  \end{tabular}
\end{table}

\section{Discussion}
The results align qualitatively with digital-humanities narratives: (1) erosion of transportation in British penal practice; (2) professionalisation of judges and a move from status-based justice to codified procedure; (3) medicalisation of madness and insanity. These shifts also coincide with known reform periods (transportation expansion and debate in the eighteenth century, penal-servitude reforms in the mid-nineteenth century, and Victorian-era medicalisation), providing lightweight event-based corroboration. A corpus-wide title filter indicates that professional honorific usage contributes to the \textit{justice} shift but does not remove it, and Global Anchors drift agrees with the SGNS ranking. After alignment-aware split-half baselines, within-bin variance is modest and net drifts remain positive across terms (Fig.~\ref{fig:netdrift}). The findings are therefore more stable than the uncorrected split-half results, but they remain hypothesis-generating without stronger external validation.

Robustness additions partially address instability: expanding $k$ reduces but does not eliminate neighborhood sparsity (Fig.~\ref{fig:overlap}); seed-variation bands suggest the mercy-retribution sign of \textit{justice} is robust (Fig.~\ref{fig:sensitivity}); corpus, anchor, and hyperparameter diagnostics guard against coverage/frequency artifacts; net-drift/z heatmaps contextualize effect sizes (Fig.~\ref{fig:netdrift}). Remaining risks include some anchoring bias despite stable-anchor checks, limited contextual baselines in sparse bins, and uncalibrated value axes (no external labels).

\section{Limitations and future work}
Open items:
\begin{itemize}
  \item Richer alignment-free usage measures (e.g., AMD/SAMD-style usage correspondence) and dependency-slot baselines to complement Global Anchors and embedding drift.
  \item Temporal chaining/reference alignments beyond the 1900s anchor, plus OT/hubness mitigation and dynamic smoothing.
  \item Historically adapted contextual models (continued pretraining on OBC, HistBERT-style) and larger context samples for APD/JSD baselines.
  \item Document-level bootstrap stability for drift and axes beyond anchor sampling; hubness diagnostics; token-based contextual baselines with balanced sampling/PRT.
  \item Calibrated value axes with external labels (e.g., sentencing outcomes or expert ratings), additional moral/procedural axes, and richer seed bootstraps.
  \item Stronger polysemy/context-diversity controls following \citep{kutuzov2022}.
\end{itemize}
Evaluating semantic change remains challenging without gold standards \citep{tahmasebi2021}; crowdsourced historical judgments or links to sentencing outcomes would strengthen validation and the “expert system” claims.

\section{Conclusion}
We demonstrated an auditable semantic-drift pipeline that couples diachronic embeddings with digital-humanities storytelling. Each module is an inference component whose outputs can be inspected and recombined. With alignment-aware split-half baselines, stable-anchor checks, hyperparameter perturbations, and preprocessing ablations, within-bin variance is modest and the main shifts remain robust across focal terms; external validation (contextual baselines, calibrated value axes, event-aligned benchmarks) would further strengthen claims of semantic change.
\section*{Data availability}
The Old Bailey Corpus is distributed under its own license and cannot be redistributed here. We provide derived drift tables, figures, and scripts to reconstruct results for licensed users.

\section*{Funding}
This work received no specific grant from any funding agency in the public, commercial, or not-for-profit sectors.

\section*{AI Disclosure Statement}
The author used OpenAI Codex to assist with code drafting and refactoring for the analysis pipeline. All code and results were reviewed and verified by the author.

\bibliographystyle{abbrvnat}
\bibliography{references}

\end{document}